\documentstyle[12pt,preprint]{aastex}  

\begin{document}

\title{Discovery and Characteristics of the Rapidly Rotating Active Asteroid (62412) 2000 SY178 in the Main Belt}
\author{Scott S. Sheppard\altaffilmark{1} and Chadwick Trujillo\altaffilmark{2}}

\altaffiltext{1}{Department of Terrestrial Magnetism, Carnegie Institution for Science, 5241 Broad Branch Rd. NW, Washington, DC 20015, USA, ssheppard@carnegiescience.edu}
\altaffiltext{2}{Gemini Observatory, 670 North A`ohoku Place, Hilo, HI 96720, USA}

\begin{abstract}  

We report a new active asteroid in the main belt of asteroids between
Mars and Jupiter.  Object (62412) 2000 SY178 exhibited a tail in
images collected during our survey for objects beyond the Kuiper Belt
using the Dark Energy Camera (DECam) on the CTIO 4 meter telescope.
We obtained broad-band colors of 62412 at the Magellan telescope,
which along with 62412's low albedo suggest it is a C-type asteroid.
62412's orbital dynamics and color strongly correlate with the Hygiea
family in the outer main belt, making it the first active asteroid
known in this heavily populated family.  We also find 62412 to have a
very short rotation period of $3.33\pm0.01$ hours from a double-peaked
light curve with a maximum peak-to-peak amplitude of $0.45\pm0.01$
magnitudes.  We identify 62412 as the fastest known rotator of the
Hygiea family and the nearby Themis family of similar composition,
which contains several known main belt comets.  The activity on 62412
was seen over 1 year after perihelion passage in its 5.6 year orbit.
62412 has the highest perihelion and one of the most circular orbits
known for any active asteroid.  The observed activity is probably
linked to 62412's rapid rotation, which is near the critical period
for break-up.  The fast spin rate may also change the shape and shift
material around 62412's surface, possibly exposing buried ice.
Assuming 62412 is a strengthless rubble pile, we find the density of
62412 to be around 1500 kg m$^{-3}$.

\end{abstract}

\keywords{Kuiper belt: general -- Oort Cloud -- comets: general -- minor planets, asteroids: general -- planets and satellites: formation}

\section{Introduction}

Comets are generally defined as objects that display dust and/or gas
emitting from their surfaces.  Observationally, this means detecting
either a coma or tail from the object.  The main source of the
observed activity in most comets is likely the sublimation of water
ice.  Comets are traditionally thought to originate in the Kuiper Belt
as short-period comets or in the Oort cloud as long-period comets.
Some comets could also originate in the Trojan regions of the giant
planets, but no definitive examples are known.  Elst et al. (1996)
reported activity around a main belt asteroid now called
133P/Elst-Pizarro.  Hsieh and Jewitt (2006) detailed the observed
activity around a few main belt asteroids between Mars and Jupiter.
Hsieh and Jewitt (2006) coined these objects Main Belt Comets (MBCs)
as it was thought that Main Belt Asteroids (MBAs) would not be able to
sustain any observable activity since water ice should be unstable on
their surfaces.  There are now at least 12 known objects well within
the main asteroid belt that have shown significant activity over the
past several years.  The source of this activity may be different for
the various objects and could include ice sublimation, impact ejecta
or rotational breakup (Jewitt 2012; Capria et al. 2012).  Because the
source of the observed activity in these main belt objects may be
unrelated to cometary ice sublimation, these objects are now
collectively called active asteroids.  The main belt comets are a
subset of the active asteroids in which water ice is thought to be the
source of the activity observed for the objects.  It is important to
determine whether the observed activity in a asteroid is driven by
volatile inventory or not as ices are important for understanding the
chemistry of the early solar nebula and planet formation
(Castillo-Rogez and Schmidt 2010; Schmidt and Castillo-Rogez 2012).

Impacts have been inferred for a few active asteroids, such as (596)
Scheila and P/2012 F5 (Gibbs), because of the large brightening and
quick fading indicative of an impulsive event like an impact (Jewitt
et al. 2010, 2011; Snodgrass et al. 2010; Ishiguro et al. 2011; Hsieh
et al 2012; Hainaut et al. 2012; Kim et al. 2012; Stevenson et
al. 2012; Moreno et al. 2012; Kleyna et al. 2013; Agarwal et
al. 2013).  Ice sublimation is inferred for some active asteroids,
such as 133P and 238P, because of repeated and prolonged activity only
seen near the perihelion of the objects (Hsieh et al. 2010,2011).
Newly discovered active asteroid P/2014 S4 (Gibbs) near the outer main
belt EOS family complex could also be from water ice sublimation as
activity was first seen near perihelion, very similar to the nearby
MBC P/2012 T1 (PanSTARRS) (Gibbs et al. 2014; Hsieh et al. 2013).
Though water ice is unstable at the surface of a MBA, it could survive
a few meters below the surface for billions of years (Schorghofer
2008; Prialnik and Rosenberg 2009).  To date, no direct or
spectroscopic ice detection has been made for an active asteroid, but
this is not unexpected because of the estimated low activity and low
active surface areas of these objects (Hsieh et al. 2011; O'Rourke et
al. 2013; Jewitt et al. 2014).

The known active asteroids are unlikely to have recently arrived at
their locations but have been within the main asteroid belt for
billions of years.  Thus these objects are not dynamically related to
the current short or long period comet populations (Levison et
al. 2006; Haghighipour 2009).  It is thought that these asteroids
likely formed near their present locations but it is possible some
could have originated in the outer solar system and become captured in
their current orbits very early in the solar system's history during
the planetary migration era (Levison et al. 2009; Walsh et
al. 2011,2012).  It is of interest to know whether the active
asteroids formed near their present locations or were captured from
the outer solar system as any volatile content within them may have
played an important role in the formation of the terrestrial planets,
including water delivery to Earth (Mottl et al. 2007; O'Brien et
al. 2014).

We have discovered a new active asteroid in the outer main belt of
asteroids.  (62412) 2000 SY178 joins the other twelve objects known to
show activity in the main belt and thus can help us further understand
these unusual objects.  We obtained detailed follow-up observations of
62412 after the initially observed activity from the object.

\section{Observations}

The main belt asteroid (62412) 2000 SY178 was serendipitously imaged
three times on UT March 28, 2014 at the CTIO 4m telescope with the
Dark Energy Camera (DECam) during our ongoing survey to find objects
beyond the Kuiper Belt (see Trujillo and Sheppard (2014) for details
of the survey).  DECam is a wide-field optical imager that covers
about 2.7 square degrees per image using sixty-two $2048\times4096$
CCD chips.  The pixel scale is 0.263 arcseconds per pixel.  Activity
was seen on three DECam images of 330 to 420 seconds taken using the
very wide VR filter with seeing at about 1 arcsecond Full Width at
Half Maximum (FWHM).  The faint tail was observed at a position angle
of $295\pm2$ degrees extending about 1 arcminute from the nucleus of
62412.

\begin{figure}
\epsscale{0.4}
\centerline{\includegraphics[angle=0,totalheight=0.6\textheight]{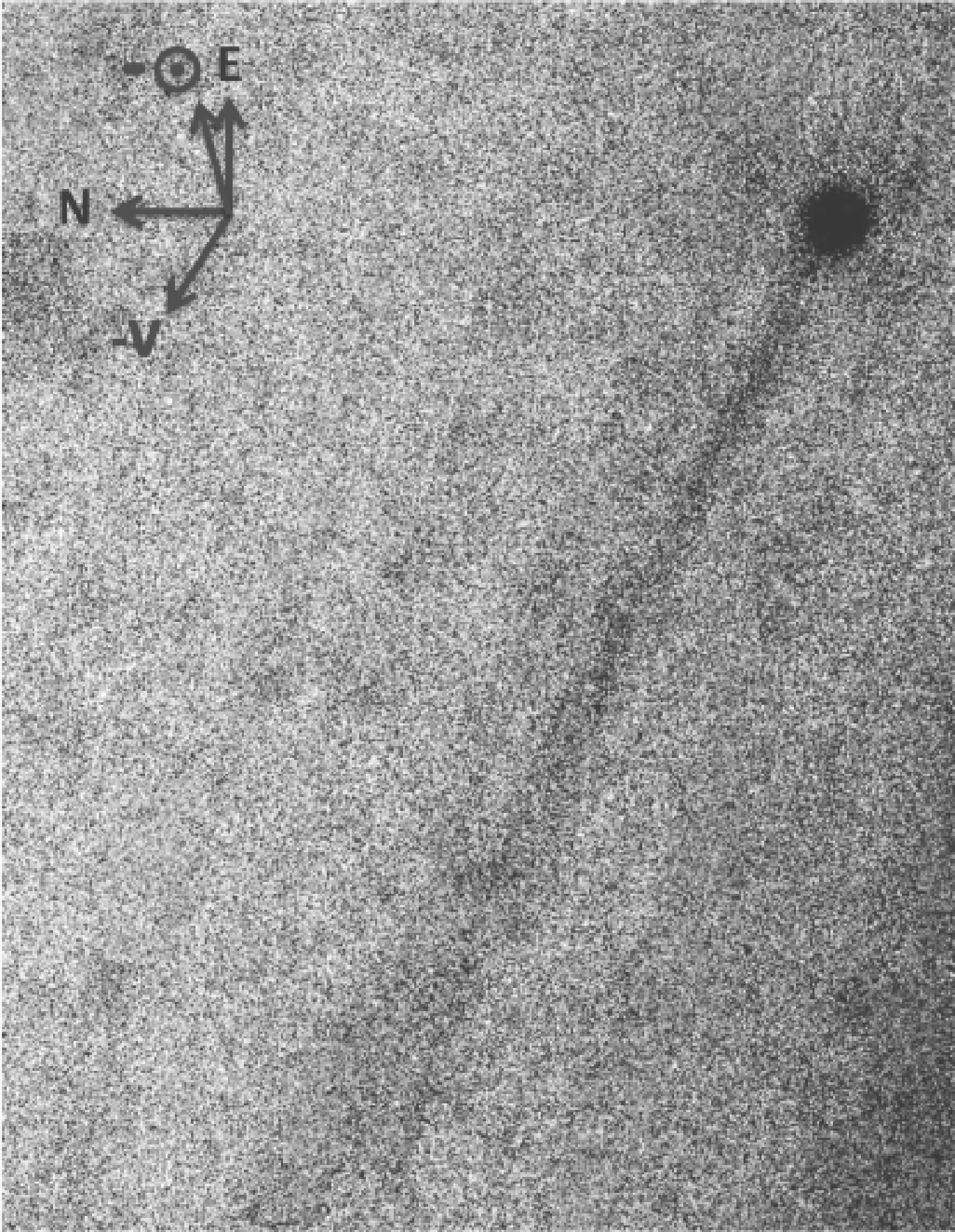}}
\caption{A 3750 second median exposure (Six 420 second VR filter and
  forty-one 30 second r'-band exposures) from IMACS on Magellan
  showing a tail for main belt object 62412.  The image is about 1.7
  arcminutes high and 1.4 arcminutes wide.  East is up and North to
  the left.  The antisolar direction ($- \sun$) and negative velocity
  vector (-V) are shown as projected on the sky.  The observed tail is
  aligned with the negative velocity vector indicating large or slow
  moving particles in the orbital plane of 62412.}
\label{fig:MBCimage} 
\end{figure}

\begin{figure}
\epsscale{0.4}
\centerline{\includegraphics[angle=0,totalheight=0.6\textheight]{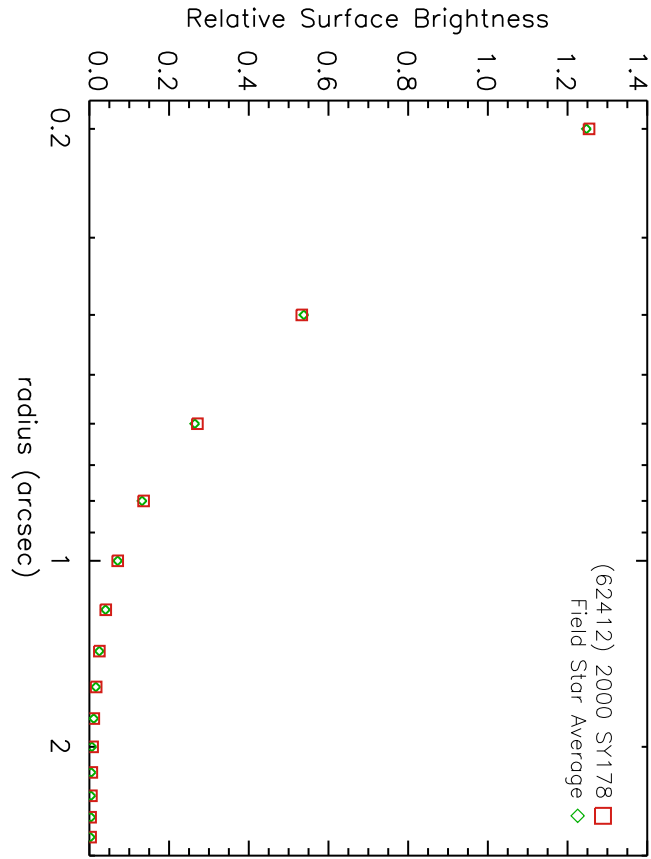}}
\caption{The relative surface brightness profile of 62412 (red
  squares) and the average of seven field stars (green diamonds) on
  the combined forty-one 30 second r'-band exposures from Magellan.
  There is no apparent extended brightness for 62412 compared to the
  field stars.  This indicates 62412 has little to no coma surrounding
  the nucleus.}
\label{fig:MBCcoma} 
\end{figure}

Follow-up observations of 62412 were obtained with the IMACS camera on
the 6.5 meter Magellan-Baade telescope on UT May 1 and 2, 2014 (Table
1).  IMACS has 8 CCD chips of $2048\times4096$ with a pixel scale of
0.20 arcseconds per pixel.  Standard image reduction techniques were
used including bias subtraction and flat-fielding from dithered
twilight images.  A clear tail for 62412 was evident at a position
angle of $297\pm 2$ degrees and extending about one arcminute from the
object in six 420 second images in the VR filter
(Figure~\ref{fig:MBCimage}).  There was no evident anti-tail on the
opposite side of the main tail of the object.  The width of the tail
is similar to the FWHM of the images suggesting it is not resolved.
Once it was confirmed that 62412 was an active asteroid it was
routinely observed using 30 second exposures over about an 8 hour time
span on May 2, 2014 in 0.8 arcsecond seeing (Table 2).  Most of these
observations were taken in the r'-band to look for variability, but
some were obtained in the g', i' and z'-bands to measure the color of
the surface of the object.  There is no obvious coma with only a faint
tail observed near 62412 and thus the photometry of the nucleus should
not be significantly contaminated (Figure~\ref{fig:MBCcoma}).  We used
an aperture of radius 3 arcseconds for photometry and the Sloan
standard star field around DLS-1359 for photometric calibration.

Four 200 second images of 62412 in the VR filter were obtain on UT
August 29, 2014 at the 8.2 meter Subaru telescope on Mauna Kea in
Hawaii.  The Suprime-Cam imager was used with a pixel scale of 0.20
arcseconds per pixel in 0.7 arcsecond seeing.  No coma or tail was
evident around 62412, suggesting the activity slowed or is no longer
present.  We also found 62412 at two epochs on archived images from
the MegaCam imager on the Canada-France-Hawaii Telescope (CFHT) using
the Solar System Object Image Search tool (SSOS) (Gwyn et al. 2012).
MegaCam has a pixel scale of 0.187 arcseconds per pixel. On UT January
25, 2012 the main image was a 384 second exposure in the r-band.
62412 was easily detected but no obvious coma or tail was seen in 1.1
arcsecond seeing, indicating the object was not active.  On UT January
4, 2013 several i-band exposures of 560 seconds each were obtain with
62412 in the field.  No visible coma or tail was observed in 0.8
arcsecond seeing in any of the i-band images, suggesting just before
perihelion the object had yet to become active.

\section{Results and Analysis}

\subsection{Orbit}

\begin{figure}
\epsscale{0.4}
\centerline{\includegraphics[angle=90,totalheight=0.6\textheight]{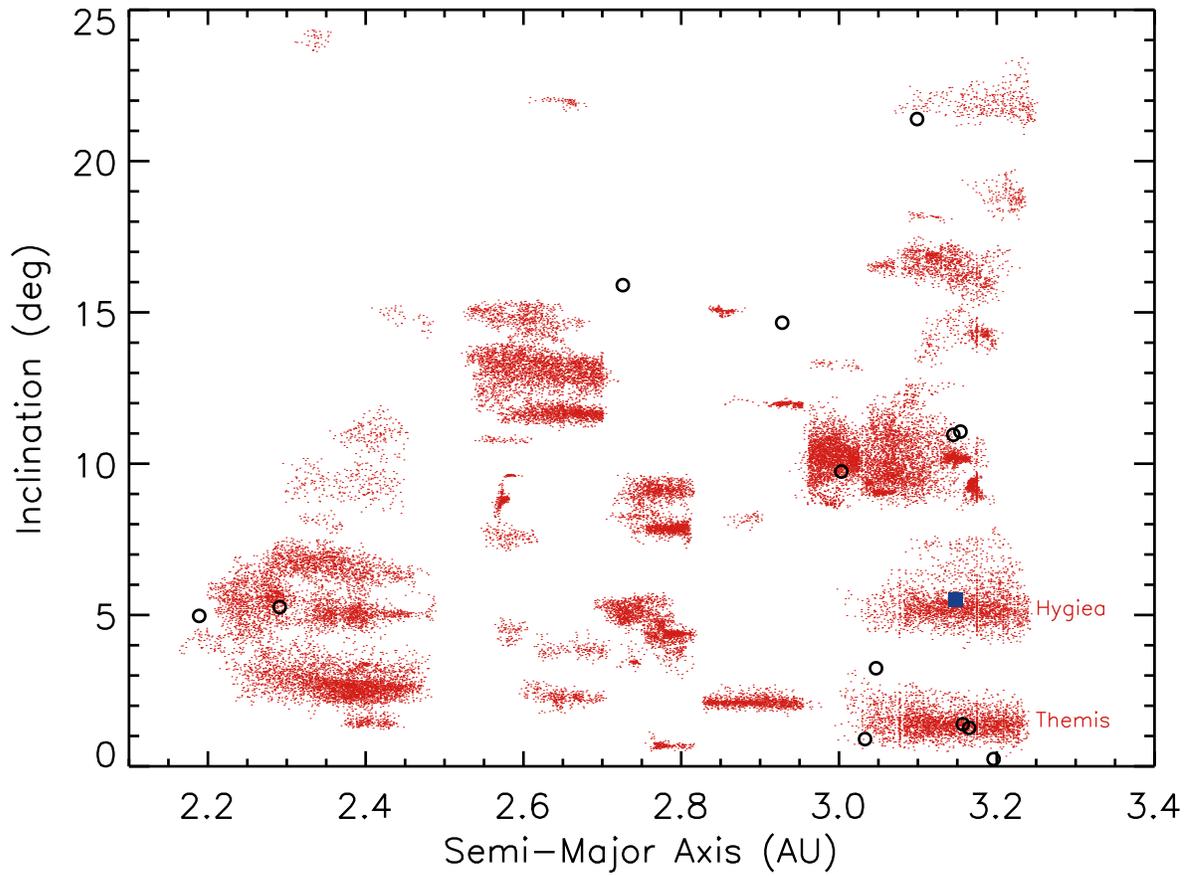}}
\caption{Semi-major axis versus inclination for known active
  asteroids (open circles), known asteroids in a family (red dots) and
  new active asteroid 62412 (solid blue square).  62412 is the first
  active asteroid known in the Hygiea family.}
\label{fig:MBCai} 
\end{figure}

Newly identified active asteroid 62412 has been observed without
activity identification for over 15 years and thus has a
well-determined orbit in the outer main belt of asteroids (Table 3).
62412 has a Tisserand parameter of 3.20, which is larger than 3 as is
typical for main belt asteroids but not short-period comets (Fernandez
et al. 2002).  Several previous active asteroids in the outer main
belt have been identified within the Themis family, but 62412's
current inclination is too high for this association
(Figure~\ref{fig:MBCai}).  In order to determine the proper orbital
elements of 62412 and look for family membership, we numerically
integrated its orbit under the influence of the planets using the
Mercury program (Chambers 1999).  We found the orbit of 62412 to be
stable for the age of the solar system with its proper orbital
elements shown in Table 3.

62412's proper inclination puts the object within the highly populated
Hygiea family (Masiero et al. 2013; Carruba 2013; Milani et al. 2014).
The Hygiea family is named after 10 Hygiea, which is a C-type asteroid
and likely the parent body of most of the family members in this
region (Parker et al. 2008; Carruba et al. 2014).  This is the first
known active asteroid within the Hygiea family.  The Hygiea family is
likely very old with an age of about 2 to 3 billion years (Nesvorny et
al. 2005; Carruba et al. 2014).

\begin{figure}
\epsscale{0.4}
\centerline{\includegraphics[angle=90,totalheight=0.6\textheight]{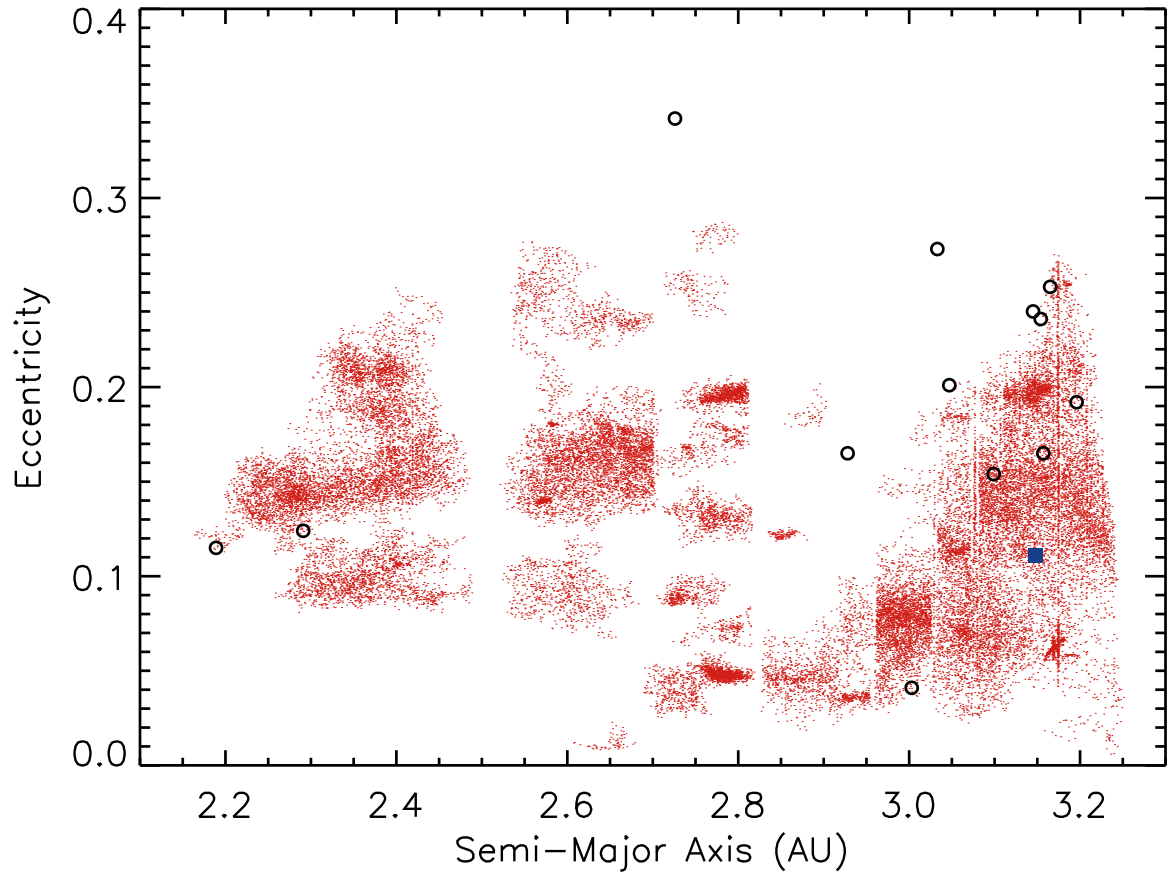}}
\caption{The semi-major axis versus eccentricity with symbols the same
  as Figure~\ref{fig:MBCai}.  62412 has one of the lowest known
  eccentricities of an active asteroid, which is currently around
  0.082 though its proper eccentricity averaged over 100 Myrs is shown
  here as 0.114 (see Table 3).}
\label{fig:MBCae}
\end{figure}

\begin{figure}
\epsscale{0.4}
\centerline{\includegraphics[angle=90,totalheight=0.6\textheight]{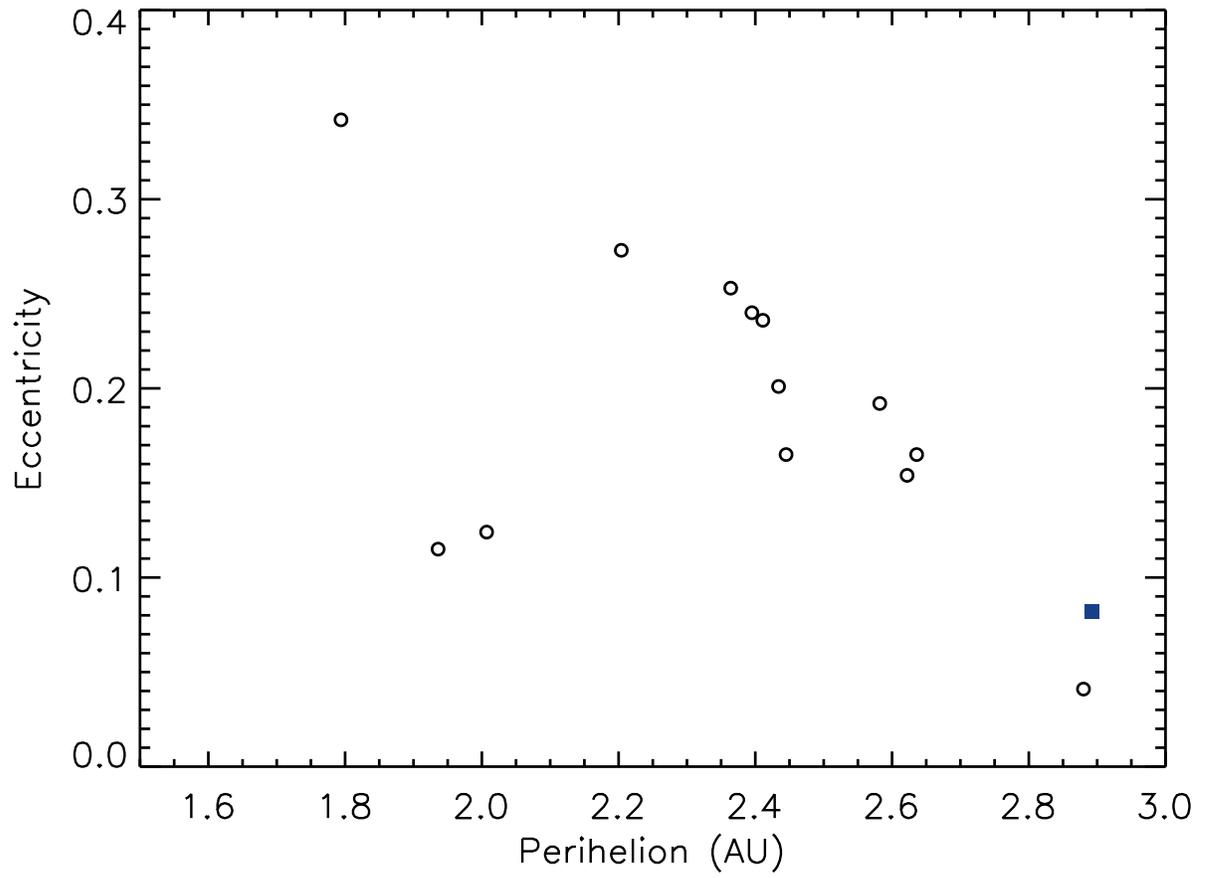}}
\caption{The perihelion versus eccentricity for known active
  asteroids.  62412 currently has the most distant perihelion and one
  of the lowest eccentricities of any known active asteroid.}
\label{fig:MBCaq}
\end{figure}

62412 passed perihelion on UT March 21, 2013 and currently has about a
5.6 year orbital period.  Thus the tail was first seen about 1 year
after perihelion, which is consistent with ice sublimation after some
thermal lag from warming near perihelion.  However, 62412 currently
has one of the lowest eccentricities and the highest perihelia of the
known active asteroids (Figures~\ref{fig:MBCae} and~\ref{fig:MBCaq}).
These orbital parameters call into question ice sublimation caused by
62412's recent perihelion passage especially compared to some other
active asteroids with closer and more eccentric orbits.  Because of
the faintness of the tail, it is currently unknown when the tail first
appeared, but it wasn't obvious in images leading up to perihelion and
appears to have faded from detection later in its orbit from our
second round of follow-up observations in late August 2014 (Table 1).
Further monitoring of 62412 is encouraged to help observe when and how
the activity arises in order to determine the cause of the activity.

\subsection{Colors, Albedo and Size}

The WISE survey serendipitously observed 62412 in May 2010 when the
object was near aphelion (Masiero et al. 2011).  Masiero et al. (2011)
found an effective radius of $5.187\pm0.171$ and optical albedo of
$0.0653\pm0.0097$ for 62412 using an H magnitude of 13.5.  This makes
62412 the second largest known active asteroid (Bauer et al. 2012).
The largest active asteroid, (596) Scheila, is believed to have
activity caused by a recent impact event (Ishiguro et al. 2011; Jewitt
et al. 2011; Moreno et al. 2011a; Hsieh et al. 2012; Bodewits et
al. 2014).

\begin{figure}
\epsscale{0.4}
\centerline{\includegraphics[angle=90,totalheight=0.6\textheight]{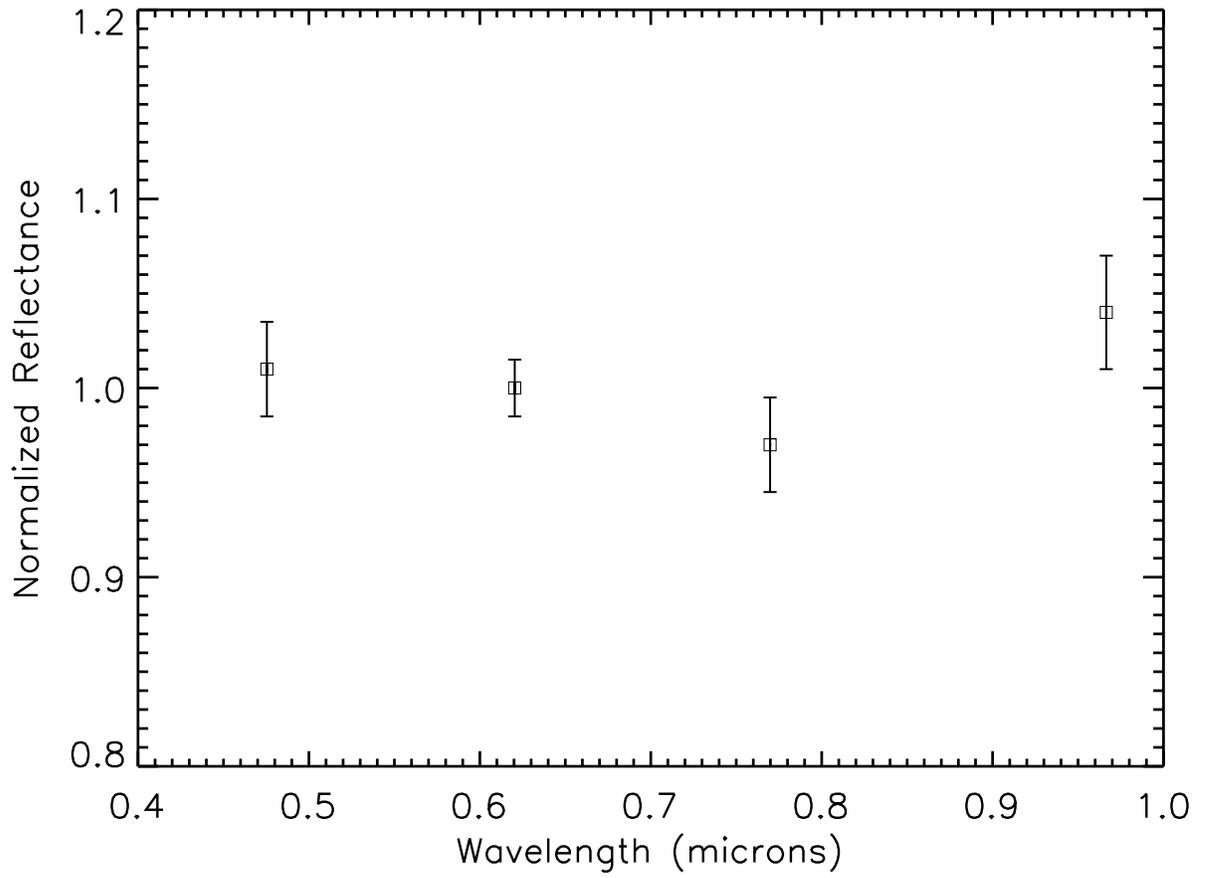}}
\caption{The normalized spectral slope of 62412 with the color of the
  Sun removed. Data are displayed at the central wavelengths of the
  g', r', i', and z' Sloan broad-band filters.  The flat spectrum is
  consistent with C-type asteroids.}
\label{fig:spectral} 
\end{figure}

The optical colors of 62412 show it to have a fairly flat spectral
slope (Table 4 and Figure~\ref{fig:spectral}).  This flat slope along
with the negative principal component color value found using the g',
r', and i' photometry as defined by Ivezic et al. (2001) and the
i'$-$z' color are consistent with a C-type asteroid (Parker et
al. 2008).  The color of 62412 correlates well with the Hygiea family
as the family is composed of mostly C-type objects (Parker et
al. 2008; Carvano et al. 2010; Carruba et al. 2014).  The low albedo
found by Masiero et al. (2011) for 62412 also suggests it is a C-type
asteroid (Mainzer et al. 2012; 2011).  These are similar
characteristics to the outer main belt MBCs such as 238P (Read) and
133P (Hsieh et al. 2004,2009).

The reduced magnitude M$_{R}(1,1,0)$ of $13.82\pm0.23$ determined for
62412 during our UT May 2, 2014 observations would be $14.18\pm0.25$
mags in the V-band, which makes the object fainter than the assumed
13.5 mag absolute magnitude used by Masiero et al. (2011) in the WISE
albedo observations.  The exact absolute magnitude depends on the
unknown phase function of 62412.  We assumed a linear 0.07 mags per
degree for the phase function based on the average of the C-type
asteroids phase curves at low phase angles (Schaefer et al. 2010).
Regardless, 62412 does not currently appear to be anomalously bright,
which suggests there is little to no coma around the object and thus
photometry should be mostly of the nucleus as shown in
Figure~\ref{fig:MBCcoma}.

The effective radius of an object can be calculated using the relation
M$_{R}(1,1,0) = m_{\odot} -
2.5\mbox{log}\left[p_{R}r_{e}^{2}/2.25\times 10^{16}\right]$ where
$m_{\odot}$ is the apparent red magnitude of the sun ($-27.07$:
Hartmann et al. 1990), $p_{R}$ is the red geometric albedo and $r_{e}$
(km) is the effective circular radius of the object.  Using the WISE
albedo of 0.0653 we get an effective radius of $3.9\pm0.3$ km, which
is smaller than found by Masiero et al. (2011) but consistent with our
fainter absolute magnitude.  But the thermal measurements from WISE
likely obtains the correct size for 62412 and it is the albedo that we
find lower, at about 0.035 (Masiero personal communication).

\subsection{Rotation}

We used the Phase Dispersion Minimization (PDM) method to determine
possible periodicity in the light curve of 62412 (Stellingwerf 1978).
In PDM, the metric is the $\Theta$ parameter, which is essentially the
variance of the unphased data divided by the variance of the data when
phased by a given period.  The best fit period should have a very
small dispersion compared to the unphased data and thus $\Theta <<$ 1
indicates that a good fit has been found.

\begin{figure}
\epsscale{0.4}
\centerline{\includegraphics[angle=90,totalheight=0.6\textheight]{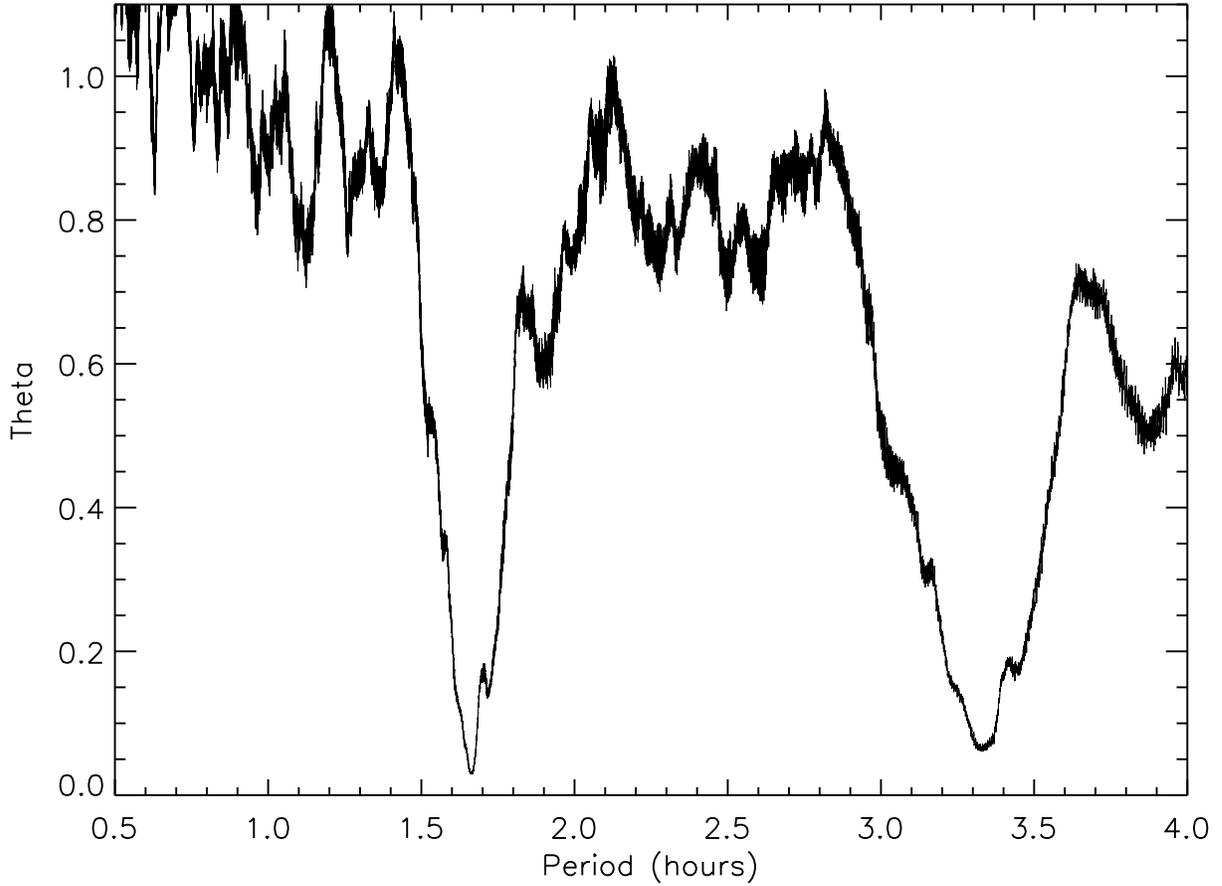}}
\caption{The Phase Dispersion Minimization (PDM) plot for 62412.  The
  best fit single-peaked period is about 1.665 hours while the best
  fit double-peaked period is about 3.33 hours.  Phasing the data
  together shows two distinct peaks and thus the double-peaked period
  at 3.33 hours is favored.}
\label{fig:MBCpdm}
\end{figure}

\begin{figure}
\epsscale{0.4}
\centerline{\includegraphics[angle=90,totalheight=0.6\textheight]{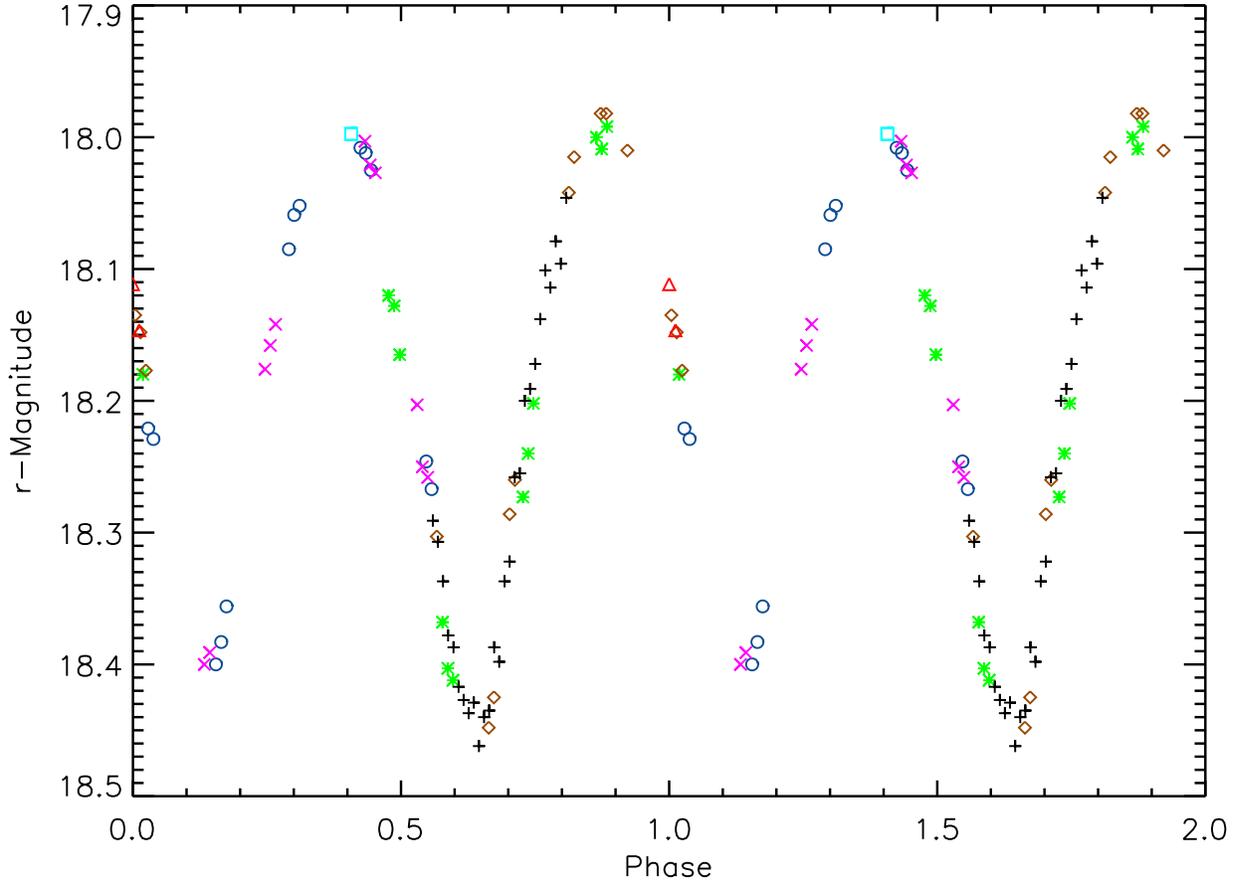}}
\caption{The phased best fit double-peaked period for 62412 of 3.33
  hours.  The maximum peak-to-peak amplitude is $0.45\pm0.01$.  The
  two lower peaks have a difference of about 0.05 magnitudes showing
  62412 has an elongated irregular shape and a light curve dominated
  by shape rather than albedo effects.  The red triangles are from May
  1, 2014 while the rest of the symbols are from May 2, 2014 where the
  different symbols each represent distinct 1.66 hour time intervals
  (each half of the double-peaked light curve).}
\label{fig:MBCphase}
\end{figure}

From Figure~\ref{fig:MBCpdm} it is apparent that a single-peaked
period of 1.665 hours and a double-peaked period of 3.33 hours are the
best fits to the light curve of 62412.  When phasing the data together
(Figure~\ref{fig:MBCphase}) it is obvious that the double-peaked
period is the best fit as the two minimum peaks have different
amplitudes.  A double-peaked periodic light curve is produced when an
elongated object's effective radius to our line of sight changes as
the object rotates.  That is, the projected cross section of an object
would go between two minima (short axis) and two maxima (long axis)
during one complete rotation.  A single-peaked light curve is likely
caused by albedo or surface variations.

Thus we find a double-peaked period of $3.33\pm0.01$ hours with a
maximum peak-to-peak amplitude of $0.45\pm0.01$ magnitudes for 62412.
The two brightest peaks, which are when the two opposite sides of
maximum surface area of the object are visible, appear to be similar.
The two fainter peaks of the light curve are different by about 0.05
magnitudes, showing the object has an elongated irregular shape.

62412 is small enough that its spin is not likely primordial and thus
has been modified over the age of the solar system (Steinberg and Sari
2014).  The large size and distant orbit of 62412 make it less
sensitive to the YORP spin-up effect than the other mostly smaller
active asteroids (Nesvorny and Vokrouhlicky 2007; Marzari et al. 2011;
Jacobson et al. 2014).  The size of 62412 is very near the transition
radius where YORP is expected to significantly modify the spins of
asteroids (Bottke et al. 2002; Pravec et al. 2002; Carbognani 2011;
Steinberg and Sari 2011).  Thus the short rotation period of 62412
could be from the YORP effect, but if not from YORP, an impact and/or
sublimation of ice off the surface could have initially spun up 62412
to the state it is now (Wiegert 2014; Polishook 2014).

\subsection{Shape, Critical Period and Density}
For an object that is elongated, the peak-to-peak amplitude of the
rotational light curve allows for the determination of the projection
of the body shape into the plane of the sky.  Assuming we are viewing
the object equatorially, the lower limit on the $a$ to $b$ axis ratio
is $a/b=10^{0.4 \Delta m}$ (Binzel et al. 1989), where $a \geq b \geq
c$ are the semi-axes with the object in rotation about the $c$ axis
and $\Delta m$ is expressed in magnitudes.  If the light curve
amplitude of 62412 is caused by an elongated shape, 62412 has $a/b
\geq 1.51$.  Using the effective radius of 5.2 km found earlier, we
get $a\sim 6.4$ km and $b\sim 4.2$ km.

An object will be near breakup if it has a rotation period near the
critical rotation period ($P_{crit}$) at which centripetal
acceleration equals gravitational acceleration towards the center of a
rotating spherical object,
\begin{equation}
P_{crit} = \left(\frac{3\pi }{G \rho}\right)^{1/2}   \label{eq:equil}
\end{equation}
where $G$ is the gravitational constant and $\rho$ is the density of
the object.  With $\rho$ = $2000$ kg m$^{-3}$, which is on the high
end of C-types but about the density of 10 Hygiea and Ceres (Baer et
al. 2011; Carry 2012), the critical period is about 2.3 hours.  At
shorter periods the object may break apart while slightly longer
periods the equilibrium figures are triaxial ellipsoids which are
elongated from the large centripetal force (Weidenschilling 1981;
Holsapple 2001).

For an elongated strengthless object, the critical period of break-up
will happen at a slower spin period (Samarasinha et al. 2004, Jewitt
2012):
\begin{equation}
P_{crit} =  (a/b \left(\frac{3\pi }{G \rho}\right))^{1/2}   \label{eq:equil}
\end{equation}
For 62412, $a/b \geq 1.51$ and thus $P_{crit} \geq 3$ hours for a
density of $2000$ kg m$^{-3}$.  Thus 62412's period appears to be
near the critical period for break-up and thus the object
should be unstable to rotational fission, especially if its density is
less than $2000$ kg m$^{-3}$, as expected for most C-type asteroids
(Mainzer et al. 2012).  Holsapple (2004) found that the spin period
for instability should be even slower than the simple discussion
above, making it even more likely that 62412 is spinning faster than
the critical period, though unknown cohesive forces will be important
(Richardson et al. 2009; Chang et al. 2014; Rozitis et al. 2014).

Assuming 62412 is at P$_{crit}$ and is a strengthless rubble pile, we
find from Equation~\ref{eq:equil} that the density would be about 1500
kg m$^{-3}$, which is significantly greater than expected for comets
(Thomas et al. 2013) and as expected for C-type asteroids of 62412's
size (Baer et al. 2011, Carry 2012).

\section{Discussion}

\subsection{Cause of 62412's Tail}
Above we show that 62412 is likely spinning near or faster than its
critical speed for breakup.  This strongly suggests that the tail
observed for 62412 is in part created by its rapid rotation.  The
observed tail is aligned with the negative velocity vector, which
usually means relatively large particles in the orbital plane (Finson
and Probstein 1968; Lisse et al. 2004; Hsieh et al. 2004; Moreno et
al. 2011b).  These large particles could be the remnants of an earlier
outburst or slow moving ejected material from 62412's rapid rotation
(see Jewitt et al. 2010; Snodgrass et al. 2010).

\begin{figure}
\epsscale{0.4}
\centerline{\includegraphics[angle=90,totalheight=0.6\textheight]{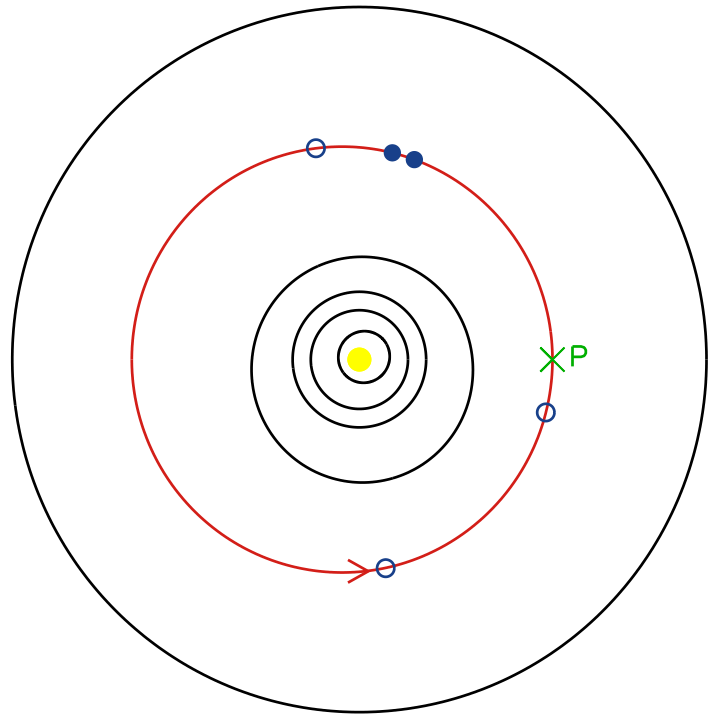}}
\caption{The orbit of 62412 (red) compared to the inner planets
  Mercury, Venus, Earth and Mars and the outer planet Jupiter (black).
  The perihelion of 62412 is shown by the green X and P while epochs
  of observations are shown by blue circles.  Filled blue circles show
  where a tail was observed and open blue circles show where the
  tail was not observed near 62412.}
\label{fig:MBCorbit}
\end{figure}

Some of the active asteroids appear to show activity within a few
months of perihelion, which has been attributed to water ice
sublimation (Hsieh et al. 2010,2011,2013).  For 62412 the first
detection of activity was in early 2014, over a year after perihelion
(Figure~\ref{fig:MBCorbit}).  62412 showed no activity leading up to
and just before perihelion in early 2013 from deep images similar to
ours that were serendipitously obtained at the CFHT telescope during
this time.  Thus the activity likely started or occurred around
perihelion or afterwards.  It is possible that 62412 has a high
thermal inertia since it was not active just before perihelion but is
active well beyond perihelion, suggesting heat took awhile to
penetrate to depth where ice could be.

The short rotation period may or may not be directly ejecting material
off the surface of 62412, but it has likely changed the shape and is
shifting or has shifted material around the object (Bottke et
al. 2002; Richardson et al. 2005; Minton 2008; Statler 2009; Harris et
al. 2009; Scheeres 2009; Holsapple 2010; Walsh et al. 2012;
Cotto-Figueroa et al. 2013; Richardson and Bowling 2014).  If ices are
contained within 62412, this shifting of material could expose fresh
ice to the surface, which could then sublimate away dragging material
with it.  The changing shape and shifting of material from the rapid
rotation may be an ongoing process that could constantly expose new
ice.  62412 is a good candidate to have ice buried just under its
surface since it is relatively large and the parent body 10 Hygiea was
recently shown to have a 3 micron spectral feature indicative of water
ice (Takir and Emery 2012).

Our last observations of 62412 in late August 2014 found no obvious
tail near the object.  Further monitoring of 62412 is necessary to
better determine the cause of its activity.  If the activity is seen
sporadically throughout 62412's orbit, fissioning of material off the
surface is most likely.  If the activity only appears near or just
after perihelion, ice sublimation would be the most likely cause,
though helped by the short rotation period.  If no further activity is
observed, it is possible 62412 experienced a recent impact that lifted
material from the surface.

\subsection{Comparison to Other Active Asteroids and Comets}
Rapid rotation has been suggested as the cause for the multiple tail
features observed for active asteroid P/2013 P5 PanSTARRS, though no
rotation period has been measured to confirm this (Jewitt et al. 2013;
Hainaut et al. 2014; Moreno et al. 2014).  Active asteroids P/2012 A2
LINEAR and P/2013 R3 Catalina-PanSTARRS have also been suggested as
possible rotationally unstable objects, though again, no rotation
periods are known for these objects and impact generated activity is a
more likely scenario for P/2012 A2 LINEAR (Jewitt et al. 2010, 2011;
Snodgrass et al. 2010; Hainaut et al. 2012; Hirabayashi et al. 2014).
Asteroid pairs, which are asteroids not bound to each other but on
very similar orbits with very similar colors could also have formed
from rotational-fission or impacts (Vokrouhlicky and Nesvorny 2008;
Pravec and Vokrouhlicky 2009; Pravec et al. 2010; Jacobson and
Scheeres 2011; Moskovitz 2012; Jacobson et al. 2014; Wolters et
al. 2014; Polishook et al. 2014).

The Themis family member active asteroid 133P has a period of 3.471
hours with an axis ratio greater than 1.45 (Hsieh et al. 2004), which
are characteristics similar to 62412. There is no obvious coma around
62412 or 133P, indicating extremely low ejection velocities for dust
particles (Jewitt et al. 2014).  Both 62412 and 133P are very fast
rotators compared to known comets (Jorda and Gutierrez 2000;
Samarasinha et al. 2004), and because of their aspherical shapes, are
likely near or faster than the critical rotational speed for breakup.
Jewitt et al. (2014) suggest the high centripetal acceleration from
133P's rapidly rotating nucleus aids in the escape of near-surface
water ice and dust from the object.

\begin{figure}
\epsscale{0.4}
\centerline{\includegraphics[angle=90,totalheight=0.6\textheight]{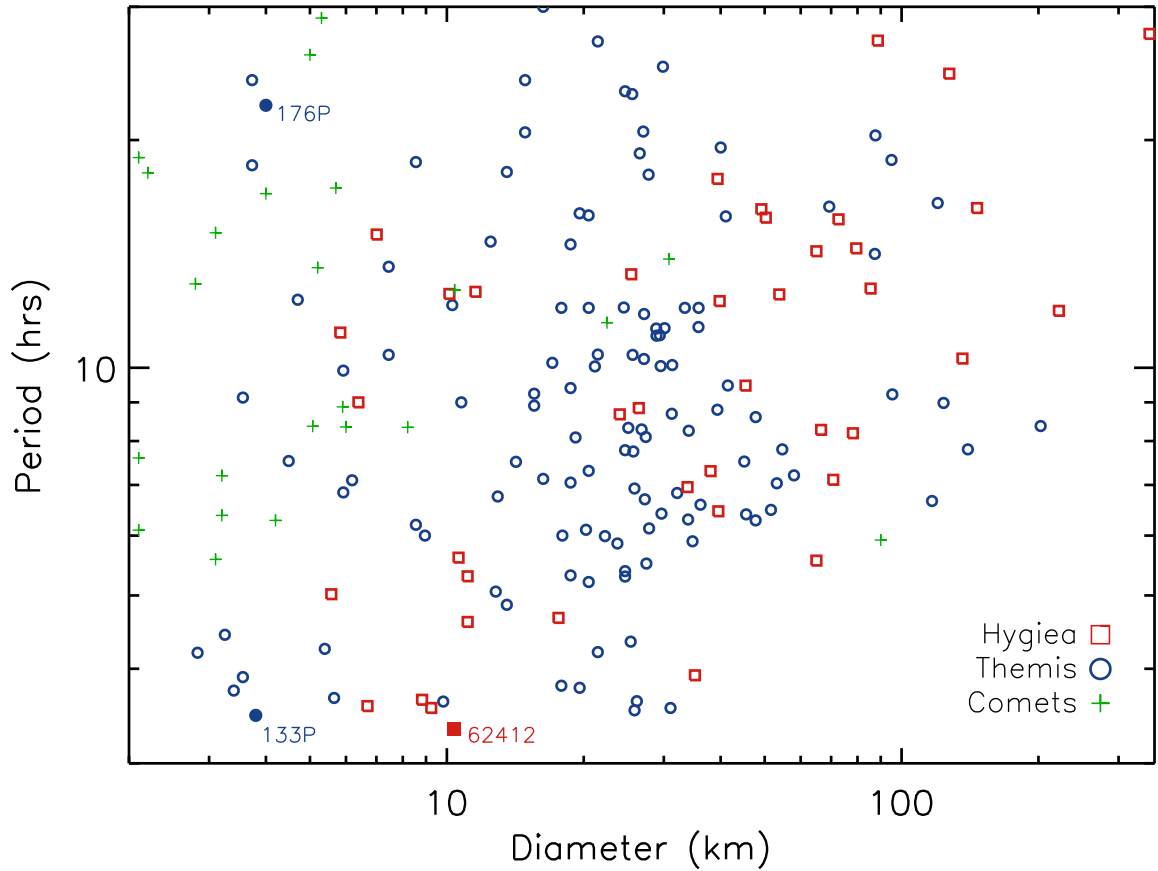}}
\caption{The size versus rotation period of asteroids in the Hygiea
  (red squares) and Themis (blue circles) asteroid families and short
  and long period comets (green pluses).  62412 has the fastest
  rotation of any member of these families (solid red square).
  Similarly, active asteroid 133P in the Themis family has the second known
  fastest period of any of these objects (solid blue circle).  This
  strongly suggests rotation is important in producing the activity of
  these objects.}
\label{fig:MBCdia_p}
\end{figure}

\begin{figure}
\epsscale{0.4}
\centerline{\includegraphics[angle=90,totalheight=0.6\textheight]{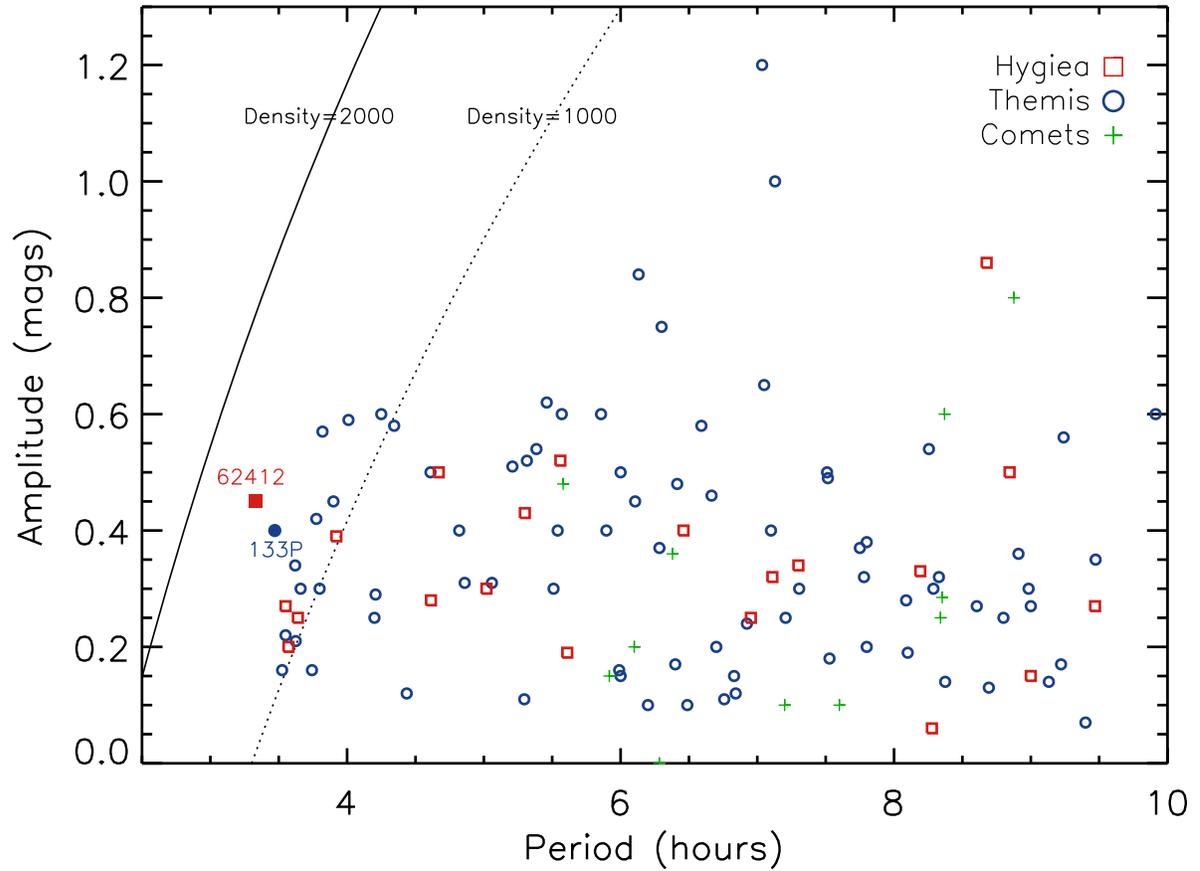}}
\caption{The periods versus the amplitudes of asteroids in the Hygiea
  (red squares) and Themis (blue circles) asteroid families and short
  and long period comets (green pluses).  The critical period for
  densities of 1000 (dotted line) and 2000 (solid line) kg m$^{-3}$
  are shown using Equation 2.  Objects to the left of these lines
  would need higher densities to be stable.  As shown in the figure,
  62412 would require the highest density to stay stable.}
\label{fig:MBCp_amp}
\end{figure}

Using data from the updated Asteroid Lightcurve Data Base (LCDB:
Warner et al. 2009), Figure~\ref{fig:MBCdia_p} shows that 62412 has
the fastest known rotation speed of any object in the mostly C-type
Hygiea and Themis families.  Active asteroid 133P has the next fastest
known rotation of any of these objects.  This strongly suggests that
these short rotation periods are an important part of the activity
seen in these objects (Figure~\ref{fig:MBCp_amp}).

62412 has a much higher density than any known short or long period
comet. The short and long period comets appear to have densities
closer to $1000$ kg m$^{-3}$, which is thought to be because of their
more distant formation in the ice rich outer solar system (Weissman et
al. 2004).  The C-type asteroids probably consist of much more rock
giving them a density closer to $2000$ kg m$^{-3}$ (Baer et al. 2011,
Carry 2012).  Thus 62412 likely formed in a much different location
than the short or long period comets.

\subsection{Number of Active Asteroids}

We have observed about 700 square degrees of sky to deeper than 24th
magnitude during our ongoing survey for objects beyond the Kuiper Belt
edge (Trujillo and Sheppard 2014).  The survey has generally been
between 5 and 20 degrees from the ecliptic.  Because of the depth and
large area, our survey has an unprecedented ability to see faint coma
and tails around objects that other surveys would not detect.  This
increased sensitivity means we can detect lower activity active
asteroids, which are also likely to be older as activity is expected
to decrease over time.  Though we would only detect the coma and/or
tail of an active object by serendipitous visual inspection, it is
generally obvious to identify activity while looking through the
survey images.  We have looked at all survey images visually while
searching for distant solar system objects.  In this way we not only
observed a tail to 62412 but also discovered comet C/2014 F3
Sheppard-Trujillo as well as detected many known comets (Sheppard and
Trujillo 2014).  The faint tail of 62412 was not detected by any other
survey even though it was within several automated surveys fields of
view and has been imaged many times in the last fifteen years.

In this crude way, we can put order of magnitude constraints on the
number of possible active asteroids in the main belt of asteroids.  We
detected about 15,000 MBAs and one active asteroid.  This detection
rate gives a ratio of active asteroid to MBA of 1:15,000.  Thus with
about 1 to 2 million main belt asteroids larger than 1 km predicted
(Tedesco et al. 2005), we would expect about 100 active asteroids,
which is consistent with earlier estimates (Hsieh 2009).

The Hygiea family likely has about 10,000 members larger than 1 km
(Tedesco et al. 2005) and thus we should only expect one as an active
asteroid.  The Themis family is a hundred times larger and thus we
would expect this family to have the majority of the active asteroids,
in which it appears this is the case as three to five of the known
active asteroids out of thirteen total are near the Themis family
(Figure~\ref{fig:MBCai}).  This of course assumes all asteroids are
just as likely to become active.  In reality the spin state,
collisional environment, surface properties and volatile content of
the asteroids are very important factors in determining what objects
become active.  This should favor more MBCs in the outer belt as they
are the most likely to have retained water ice over the age of the
solar system as shown by the recent possible detections of water ice
on Ceres, Themis and Cybele (Rivkin and Emery 2010, Campins et
al. 2010; Licandro et al. 2011; Kuppers et al. 2014).

\section{Summary}

We report the 13th known active asteroid in the main asteroid belt,
(62412) 2000 SY178.  There is no obvious coma around the object but it
exhibits a faint tail.  62412 is about 5 km in radius, making it the
second largest known active asteroid.  The main results of this paper
are:

1) The orbit of 62412 is stable for the age of the solar system and
has a Tisserand parameter greater than 3, which is typical of main
belt asteroids.  We determine that proper orbital elements for 62412
make it the first known active asteroid in the Hygiea family in the
outer belt of asteroids.

2) The optical colors of 62412 match those of typical primitive C-type
asteroids, which is the dominate type of asteroid in the Hygiea family.

3) 62412 has the highest perihelion and one of the lowest
eccentricities of the known active asteroids.  The observed tail was
first seen over 13 months after perihelion with no activity observed
just before perihelion.

4) We find a rapid rotation of $3.33\pm0.01$ hours for 62412.  The
double-peaked peak-to-peak light curve amplitude of $0.45\pm0.01$
magnitudes suggests 62412 is an elongated object with an $a$ to $b$
axial ratio greater than or equal to 1.51.

5) 62412 is likely rotating faster than the critical period for
rotational breakup of a strengthless body.  To prevent rotational
fission, it must have a density of about 1500 kg m$^{-3}$.  This is
typical for C-type asteroids and well above known short and long
period comet densities.

6) 62412 has the fastest spin rate of any known member of the Hygiea
or Themis asteroid families as well as any long or short period comet.
This suggests the spin state may be the cause or is at least a major
factor in the observed activity.  The rapid rotation may allow
particles to directly escape from the surface and likely causes
changes in the shape and shifts material around on 62412, which could
expose possible buried ices.

7) The number of active asteroids in the main belt is likely to be
around 100 objects, consistent with previous estimates.

\section*{Acknowledgments}
This project used data obtained with the Dark Energy Camera (DECam),
which was constructed by the Dark Energy Survey (DES) collaborating
institutions: Argonne National Lab, University of California Santa
Cruz, University of Cambridge, Centro de Investigaciones Energeticas,
Medioambientales y Tecnologicas-Madrid, University of Chicago,
University College London, DES-Brazil consortium, University of
Edinburgh, ETH-Zurich, University of Illinois at Urbana-Champaign,
Institut de Ciencies de l'Espai, Institut de Fisica d'Altes Energies,
Lawrence Berkeley National Lab, Ludwig-Maximilians Universitat,
University of Michigan, National Optical Astronomy Observatory,
University of Nottingham, Ohio State University, University of
Pennsylvania, University of Portsmouth, SLAC National Lab, Stanford
University, University of Sussex, and Texas A\&M University. Funding
for DES, including DECam, has been provided by the U.S. Department of
Energy, National Science Foundation, Ministry of Education and Science
(Spain), Science and Technology Facilities Council (UK), Higher
Education Funding Council (England), National Center for
Supercomputing Applications, Kavli Institute for Cosmological Physics,
Financiadora de Estudos e Projetos, Fundacao Carlos Chagas Filho de
Amparo a Pesquisa, Conselho Nacional de Desenvolvimento Cientifico e
Tecnologico and the Ministerio da Ciencia e Tecnologia (Brazil), the
German Research Foundation-sponsored cluster of excellence ``Origin
and Structure of the Universe'' and the DES collaborating
institutions.  Observations were partly obtained at Cerro Tololo
Inter-American Observatory, National Optical Astronomy Observatory,
which are operated by the Association of Universities for Research in
Astronomy, under contract with the National Science Foundation.
C.T. is supported by the Gemini observatory, which is operated by the
Association of Universities for Research in Astronomy, Inc., on behalf
of the international Gemini partnership of Argentina, Australia,
Brazil, Canada, Chile, the United Kingdom, and the United States of
America.  This research used the facilities of the Canadian Astronomy
Data Centre operated by the National Research Council of Canada with
the support of the Canadian Space Agency.  We thank H. Hsieh for
comments on this manuscript.  This paper includes data gathered with
the 6.5 meter Magellan Telescopes located at Las Campanas Observatory,
Chile.  This research was funded by NASA Planetary Astronomy grant
NNX12AG26G.

\newpage



\begin{center}
\begin{deluxetable}{lcccccccc}
\tablenum{1}
\tablewidth{6.2 in}
\tablecaption{Geometry of the 62412 Observations}
\tablecolumns{9}
\tablehead{
\colhead{UT Date} & \colhead{$R$}  & \colhead{$\Delta$} & \colhead{$\alpha$} & \colhead{Plane} & \colhead{True} & \colhead{PA$_{Sun}$} & \colhead{PA$_{v}$} & \colhead{Active} \\ \colhead{} &\colhead{(AU)} &\colhead{(AU)} & \colhead{(deg)} & \colhead{(deg)} & \colhead{(deg)} & \colhead{(deg)} & \colhead{(deg)} & \colhead{}}  
\startdata
2012 Jan 25.307       &  3.102  &  3.028  &  18.4 & -0.64 & 277.3  & 68.5  &  246.3 &  No \nl
2013 Jan 04.407       &  2.900  &  1.950  &  6.12 & -1.69 & 344.4  & 271.4 & 287.6 &  No \nl
2013 Mar 21           &  2.892  &  2.357  &  \multicolumn{6}{c}{Perihelion} \nl
2014 Mar 28.245       &  3.062  &  2.144  &  8.78 & 1.35  & 74.4   & 300.9 & 291.8 & Yes \nl
2014 May 01.192       &  3.088  &  2.101  &  4.44 & 2.15  & 80.6   & 84.9  & 293.8 & Yes \nl
2014 May 02.005       &  3.089  &  2.104  &  4.75 & 2.16  & 80.8   & 87.0  & 293.9 & Yes \nl
2014 Aug 29.246       &  3.184  &  3.534  &  16.2 & -0.15 & 101.8  & 111.7 & 291.2 & No \nl
2016 Jan 10           &  3.411  &  4.332  &  \multicolumn{6}{c}{Aphelion} \nl
\enddata
\tablenotetext{}{Quantities are the heliocentric distance ($R$), geocentric distance ($\Delta$), phase angle ($\alpha$), the orbit plane angle which is the angle between the observer and target orbital plane (Plane), true anomaly (True), position angle of the antisolar vector as projected in the plane of the sky (PA$_{Sun}$), and position angle of the negative velocity vector as projected in the plane of the sky (PA$_{v}$). UT Date shows the year, month, and time of day at the start of the observations on each night it was observed.  Individual exposure times used for colors and the lightcurve measurments on May 2, 2014 were 30 seconds in the g', r', i' and z' filters and they were rotated after each observation to prevent any rotational light curves from influencing the color calculations.}
\end{deluxetable}
\end{center}


\newpage



\begin{center}
\begin{deluxetable}{lcccc}
\tablenum{2}
\tablewidth{6.5 in}
\tablecaption{r'-band Observations of (62412) 2000 SY178 \label{tab:obs62412}}
\tablecolumns{5}
\tablehead{
\colhead{Airmass} & \colhead{Exp\tablenotemark{b}}  & \colhead{UT Date\tablenotemark{c}} & \colhead{Mag.\tablenotemark{d}}  & \colhead{Err\tablenotemark{e}} \\  \colhead{} & \colhead{(sec)} & \colhead{yyyy mm dd.ddddd}  & \colhead{($m_{r'}$)} & \colhead{($m_{r'}$)} }
\startdata
1.239 &  45 &  2014 05 01.257820 &   18.11  &  0.01  \nl 
1.247 &  30 &  2014 05 01.259440 &   18.15  &  0.01  \nl 
1.524 &  35 &  2014 05 02.008044 &   18.00  &  0.01  \nl 
1.436 &  35 &  2014 05 02.017719 &   18.12  &  0.01  \nl 
1.424 &  35 &  2014 05 02.019155 &   18.13  &  0.01  \nl 
1.413 &  35 &  2014 05 02.020601 &   18.17  &  0.01  \nl 
1.335 &  30 &  2014 05 02.031678 &   18.37  &  0.01  \nl 
1.326 &  30 &  2014 05 02.033055 &   18.40  &  0.01  \nl 
1.318 &  30 &  2014 05 02.034398 &   18.41  &  0.01  \nl 
1.221 &  30 &  2014 05 02.052500 &   18.27  &  0.01  \nl 
1.215 &  30 &  2014 05 02.053842 &   18.24  &  0.01  \nl 
1.209 &  30 &  2014 05 02.055208 &   18.20  &  0.01  \nl 
1.146 &  30 &  2014 05 02.071504 &   18.00  &  0.01  \nl 
1.142 &  30 &  2014 05 02.072858 &   18.01  &  0.01  \nl 
1.137 &  30 &  2014 05 02.074213 &   17.99  &  0.01  \nl 
1.087 &  30 &  2014 05 02.092905 &   18.18  &  0.01  \nl 
1.084 &  30 &  2014 05 02.094259 &   18.22  &  0.01  \nl 
1.081 &  30 &  2014 05 02.095613 &   18.23  &  0.01  \nl 
1.052 &  30 &  2014 05 02.111794 &   18.40  &  0.01  \nl 
1.050 &  30 &  2014 05 02.113160 &   18.38  &  0.01  \nl 
1.048 &  30 &  2014 05 02.114525 &   18.36  &  0.01  \nl 
1.031 &  30 &  2014 05 02.130683 &   18.09  &  0.01  \nl 
1.030 &  30 &  2014 05 02.132049 &   18.06  &  0.01  \nl 
1.029 &  30 &  2014 05 02.133437 &   18.05  &  0.01  \nl 
1.024 &  30 &  2014 05 02.149190 &   18.01  &  0.01  \nl 
1.024 &  30 &  2014 05 02.150544 &   18.01  &  0.01  \nl 
1.024 &  30 &  2014 05 02.151910 &   18.03  &  0.01  \nl 
1.027 &  30 &  2014 05 02.166215 &   18.25  &  0.01  \nl 
1.028 &  30 &  2014 05 02.167593 &   18.27  &  0.01  \nl 
1.029 &  30 &  2014 05 02.168958 &   18.30  &  0.01  \nl 
1.040 &  30 &  2014 05 02.182384 &   18.45  &  0.01  \nl 
1.042 &  30 &  2014 05 02.183727 &   18.43  &  0.01  \nl 
1.047 &  30 &  2014 05 02.187801 &   18.29  &  0.01  \nl 
1.049 &  30 &  2014 05 02.189167 &   18.26  &  0.01  \nl 
1.072 &  30 &  2014 05 02.203125 &   18.04  &  0.01  \nl 
1.074 &  30 &  2014 05 02.204479 &   18.02  &  0.01  \nl 
1.089 &  30 &  2014 05 02.211343 &   17.98  &  0.01  \nl 
1.092 &  30 &  2014 05 02.212778 &   17.98  &  0.01  \nl 
1.106 &  30 &  2014 05 02.218264 &   18.01  &  0.01  \nl 
1.139 &  30 &  2014 05 02.229653 &   18.14  &  0.01  \nl 
1.144 &  30 &  2014 05 02.231030 &   18.15  &  0.01  \nl 
1.148 &  30 &  2014 05 02.232419 &   18.18  &  0.01  \nl 
1.207 &  30 &  2014 05 02.247558 &   18.40  &  0.01  \nl 
1.213 &  30 &  2014 05 02.248935 &   18.39  &  0.01  \nl 
1.285 &  30 &  2014 05 02.263252 &   18.18  &  0.01  \nl 
1.293 &  30 &  2014 05 02.264641 &   18.16  &  0.01  \nl 
1.301 &  30 &  2014 05 02.265995 &   18.14  &  0.01  \nl 
1.466 &  30 &  2014 05 02.289132 &   18.00  &  0.01  \nl 
1.478 &  30 &  2014 05 02.290486 &   18.02  &  0.01  \nl 
1.490 &  30 &  2014 05 02.291829 &   18.03  &  0.01  \nl 
1.598 &  30 &  2014 05 02.302627 &   18.20  &  0.01  \nl 
1.613 &  30 &  2014 05 02.303981 &   18.25  &  0.01  \nl 
1.628 &  30 &  2014 05 02.305370 &   18.26  &  0.01  \nl 
1.644 &  30 &  2014 05 02.306678 &   18.29  &  0.01  \nl 
1.659 &  30 &  2014 05 02.307998 &   18.31  &  0.01  \nl 
1.675 &  30 &  2014 05 02.309306 &   18.34  &  0.01  \nl 
1.692 &  30 &  2014 05 02.310625 &   18.38  &  0.01  \nl 
1.710 &  30 &  2014 05 02.312014 &   18.39  &  0.01  \nl 
1.727 &  30 &  2014 05 02.313333 &   18.42  &  0.01  \nl 
1.745 &  30 &  2014 05 02.314641 &   18.43  &  0.01  \nl 
1.763 &  30 &  2014 05 02.315949 &   18.44  &  0.01  \nl 
1.782 &  30 &  2014 05 02.317269 &   18.43  &  0.01  \nl 
1.802 &  30 &  2014 05 02.318611 &   18.46  &  0.01  \nl 
1.821 &  30 &  2014 05 02.319919 &   18.44  &  0.01  \nl 
1.842 &  30 &  2014 05 02.321238 &   18.44  &  0.01  \nl 
1.862 &  30 &  2014 05 02.322546 &   18.39  &  0.01  \nl 
1.884 &  30 &  2014 05 02.323854 &   18.40  &  0.01  \nl 
1.907 &  30 &  2014 05 02.325231 &   18.34  &  0.01  \nl 
1.929 &  30 &  2014 05 02.326539 &   18.32  &  0.01  \nl 
1.953 &  30 &  2014 05 02.327859 &   18.26  &  0.01  \nl 
1.977 &  30 &  2014 05 02.329167 &   18.26  &  0.01  \nl 
2.001 &  30 &  2014 05 02.330475 &   18.20  &  0.01  \nl 
2.027 &  30 &  2014 05 02.331840 &   18.19  &  0.01  \nl 
2.054 &  30 &  2014 05 02.333160 &   18.17  &  0.01  \nl 
2.080 &  30 &  2014 05 02.334468 &   18.14  &  0.01  \nl 
2.108 &  30 &  2014 05 02.335775 &   18.10  &  0.01  \nl 
2.137 &  30 &  2014 05 02.337095 &   18.11  &  0.01  \nl 
2.168 &  30 &  2014 05 02.338472 &   18.08  &  0.01  \nl 
2.198 &  30 &  2014 05 02.339792 &   18.10  &  0.01  \nl 
2.232 &  30 &  2014 05 02.341192 &   18.05  &  0.01  \nl 
 \enddata
\tablenotetext{b}{Exposure time for the image.}
\tablenotetext{c}{Decimal Universal Time at the start of the integration.}
\tablenotetext{d}{Apparent magnitude in r' filter.}
\tablenotetext{e}{Uncertainties on the individual photometric measurements.}
\end{deluxetable}
\end{center}


\newpage



\begin{center}
\begin{deluxetable}{lccccccc}
\tablenum{3}
\tablewidth{6 in}
\tablecaption{Orbital Information for 62412}
\tablecolumns{8}
\tablehead{
\colhead{Type} & \colhead{$a$} & \colhead{$e$}  & \colhead{$i$} & \colhead{$q$} & \colhead{$Q$} & \colhead{$P$} & \colhead{$T_{J}$} \\ \colhead{} & \colhead{(AU)} & \colhead{}  & \colhead{(deg)} &\colhead{(AU)} &\colhead{(AU)} & \colhead{(yr)} & \colhead{} }  
\startdata
Current        & 3.151   & 0.082  &  4.74 & 2.892 & 3.410 & 5.6   &  3.20  \nl   
Proper         & 3.147   & 0.114  &  5.64 & 2.790 & 3.506 & 5.6   &  3.19   \nl
\enddata
\tablenotetext{}{Quantities are the semi-major axis ($a$), eccentricity ($e$), inclination ($i$), perihelion distance ($q$), aphelion distance ($Q$), period ($P$), and Tisserand parameter with respect to Jupiter ($T_{J}$).  Current orbital elements are from the Minor Planet Center. Proper orbital elements are from our 4.6 billion year numerical integration using the Mercury code (Chambers 1999).} 
\end{deluxetable}
\end{center}


\newpage



\begin{center}
\begin{deluxetable}{lcccc}
\tablenum{4}
\tablewidth{6.0 in}
\tablecaption{Sloan g',r',i',z' Optical Photometry}
\tablecolumns{5}
\tablehead{
\colhead{$m_{r'}$} & \colhead{M$_{R}(1,1,0)$} & \colhead{g'-r'} & \colhead{r'-i'} & \colhead{i'-z'} \\ \colhead{(mag)} & \colhead{} & \colhead{(mag)} & \colhead{(mag)} & \colhead{(mag)} }  
\startdata
$18.215\pm0.225$   & $13.82\pm0.01$  &  $0.43\pm0.02$ &  $0.14\pm0.02$ &  $-0.01\pm0.02$ \nl
\enddata
\tablenotetext{}{The Sloan colors converted to BVRI using Smith et al. (2002) are $R=18.02\pm0.01$ with $B-R=1.00\pm0.02$, $V-R=0.36\pm0.02$, and $R-I=0.35\pm0.02$.  The error on $m_{r'}$ is large because of the significant light curve while the error on M$_{R}(1,1,0)$ is shown to be much smaller as we use the mean of the light curve.}
\end{deluxetable}
\end{center}


\newpage

\newpage

\newpage

\newpage

\newpage

\newpage

\newpage

\newpage

\newpage

\newpage

\newpage

\end{document}